\documentclass[aps,prc,twocolumn,nofootinbib,amsmath,showpacs,superscriptaddress,floatfix]{revtex4}
\usepackage{graphicx,epsfig,epstopdf,longtable}
\usepackage{mathptmx}
\usepackage[light,first]{draftcopy} 
\usepackage[pdftex]{hyperref}
\usepackage{wasysym}

\newcommand{\beqy}{\begin{eqnarray}}
\newcommand{\eeqy}{\end{eqnarray}}
\newcommand{\bmlet}{\begin{subequations}}
\newcommand{\emlet}{\end{subequations}}
\newcounter{saveeqn}

\def\gsimeq{\,\,\raise0.14em\hbox{$>$}\kern-0.76em\lower0.28em\hbox  
{$\sim$}\,\,}  
\def\lsimeq{\,\,\raise0.14em\hbox{$<$}\kern-0.76em\lower0.28em\hbox  
{$\sim$}\,\,}  

\begin{document}

\title{Constant-temperature level densities in the quasi-continuum of Th and U isotopes}

\author{M.~Guttormsen}
\email{magne.guttormsen@fys.uio.no}
\affiliation{Department of Physics, University of Oslo, N-0316 Oslo, Norway}
\author{B.~Jurado}
\affiliation{CENBG, CNRS/IN2P3, UniversitŽ Bordeaux I, Chemin du Solarium B.P.~120, 33175 Gradignan, France}
\author{J.N.~Wilson}
\affiliation{Institut de Physique Nucleaire d'Orsay, Bat. 100, 15 rue G. Glemenceau, 91406 Orsay Cedex, France}
\author{M.~Aiche}
\affiliation{CENBG, CNRS/IN2P3, UniversitŽ Bordeaux I, Chemin du Solarium B.P.~120, 33175 Gradignan, France}
\author{L.A.~Bernstein}
\affiliation{Lawrence Livermore National Laboratory, 7000 East Avenue, Livermore, CA 94550-9234, USA}
\author{Q.~Ducasse}
\affiliation{CENBG, CNRS/IN2P3, UniversitŽ Bordeaux I, Chemin du Solarium B.P.~120, 33175 Gradignan, France}
\author{F.~Giacoppo}
\affiliation{Department of Physics, University of Oslo, N-0316 Oslo, Norway}
\author{A.~G{\"o}rgen}
\affiliation{Department of Physics, University of Oslo, N-0316 Oslo, Norway}
\author{F.~Gunsing}
\affiliation{CEA Saclay, DSM/Irfu/SPhN, F-91191 Gif-sur-Yvette Cedex, France}
\author{T.W.~Hagen}
\affiliation{Department of Physics, University of Oslo, N-0316 Oslo, Norway}
\author{A.C.~Larsen}
\affiliation{Department of Physics, University of Oslo, N-0316 Oslo, Norway}
\author{M.~Lebois}
\affiliation{Institut de Physique Nucleaire d'Orsay, Bat. 100, 15 rue G. Glemenceau, 91406 Orsay Cedex, France}
\author{B.~Leniau}
\affiliation{Institut de Physique Nucleaire d'Orsay, Bat. 100, 15 rue G. Glemenceau, 91406 Orsay Cedex, France}
\author{T.~Renstr{\o}m}
\affiliation{Department of Physics, University of Oslo, N-0316 Oslo, Norway}
\author{S.J.~Rose}
\affiliation{Department of Physics, University of Oslo, N-0316 Oslo, Norway}
\author{S.~Siem}
\affiliation{Department of Physics, University of Oslo, N-0316 Oslo, Norway}
\author{T.~Tornyi}
\affiliation{Department of Physics, University of Oslo, N-0316 Oslo, Norway}
\author{G.M.~Tveten}
\affiliation{Department of Physics, University of Oslo, N-0316 Oslo, Norway}
\author{M.~Wiedeking}
\affiliation{iThemba LABS, P.O. Box 722, 7129 Somerset West, South Africa}
\date{\today}

\begin{abstract}
Particle-$\gamma$ coincidences have been measured to obtain $\gamma$-ray spectra as a function of excitation energy for $^{231-233}$Th and $^{237-239}$U. The level densities, which were extracted using the Oslo method, show a constant temperature behavior. The isotopes display very similar temperatures in the quasi-continuum, however, the even-odd isotopes reveal a constant entropy increase $\Delta S$ compared to their even-even neighbors. The entropy excess depends on available orbitals for the last unpaired valence neutron of the heated nuclear system. Also, experimental microcanonical temperature and heat capacity have been extracted. Several poles in the heat capacity curve support the idea that an almost continuous melting of Cooper pairs is responsible for the constant-temperature behavior.
\end{abstract}

\pacs{21.10.Ma, 25.20.Lj, 27.90.+b, 25.40.Hs}

\maketitle

\section{Introduction}
\label{sec:int}
The knowledge of level density in the actinides is of great importance for modeling nuclear reactions used in fuel-cycle calculations of fast nuclear reactors. In addition, it has the potential of improving the nuclear-physics aspect of the nucleosynthesis for the heaviest nuclear systems in astrophysical environments.

The first theoretical attempt to describe nuclear level densities was performed by Hans Bethe in 1936~\cite{bethe36}. In this pioneering work, the nucleus was described as a gas of non-interacting fermions moving in equally spaced single-particle orbitals. The Fermi-gas model was later refined by introducing a shift in the excitation energy $E$ that takes into account the increase of the ground-state binding energy due to pairing correlations. This parameterization of the Fermi-gas model has been popular for many decades~\cite{capote2009}.
 
A characteristic property of the Fermi-gas model is that the nuclear temperature follows a $T \propto \sqrt{E}$ dependency. However, as more and more data have become available in the quasi-continuum region, there is less support for the Fermi-gas model. Experimental results using the Oslo method~\cite{guttormsen2003} and particle evaporation techniques~\cite{voinov2009} support the constant temperature picture. Typically, the temperature is found to be constant above $ E \approx 2\Delta$, where $\Delta$ is the pairing gap parameter.

From several nuclear-level density studies using the Oslo method, it seems that the constant temperature description works best for heavier well-deformed systems. To follow this trend further, we study for the first time the level densities in the quasi-continuum of actinides, where one expects a uniform and dense occurrence of single-particle orbitals.

The actinides have a very high level density of several million levels per MeV already at 5 MeV of excitation energy. The close-lying nature of these levels makes it impossible to detect all of them with conventional spectroscopy;  in some cases the level density can only be determined up to a few 100 keV of excitation energy from the counting of low-lying discrete known levels~\cite{ENSDF}. At the neutron separation energy $S_n$, there is reliable level-density information from neutron resonances; however, this information is restricted in energy as well as in spin range. Between the discrete levels and the separation energy, we are not aware of any data in the literature that provide further information on the level density of the actinides.

The Oslo nuclear physics group has developed a method~\cite{Schiller00,Lars11} to determine simultaneously the level density and the $\gamma$-ray strength function ($\gamma$SF) from particle-$\gamma$ coincidences. In this work, the Oslo method is applied to extract the level densities of the $^{231-233}$Th and $^{237-239}$U isotopes. Recently~\cite{guttormsen2012}, the $\gamma$SFs in $^{231-233}$Th and $^{232,233}$Pa were reported. 

Section II describes the experimental techniques and methods, and in Sec.~III the extraction and normalization of the level densities are discussed. In Sec.~IV the thermodynamic aspects of the actinides are studied, and the conclusions are drawn in Sec.~V.

\section{Experiments}
\label{sec:exp}

Two experiments with targets $^{232}$Th and $^{238}$U were conducted at the Oslo Cyclotron Laboratory (OCL). The selfsupporting $^{232}$Th target (thickness 0.968 mg/cm$^2$) was bombarded with a 12-MeV deuteron and a 24-MeV $^{3}$He beam. The $^{238}$U target (thickness 0.250 mg/cm$^2$ and enrichment 99.7\%) had a carbon backing (thickness 0.040 mg/cm$^2$) and was bombarded with a 15-MeV deuteron beam. Particle-$\gamma$ coincidences were measured with the SiRi particle telescope and the CACTUS $\gamma$-detector system~\cite{siri,CACTUS}.
 \begin{figure*}[t]
 \begin{center}
 \includegraphics[clip,width=2\columnwidth]{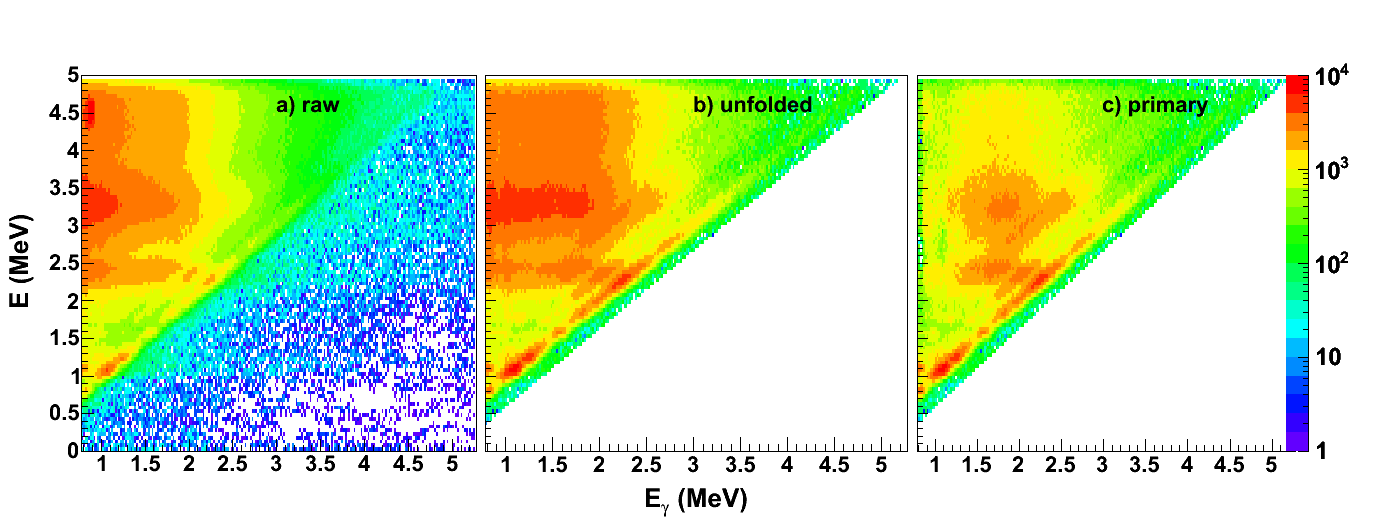}
 \caption{(Color online) Initial excitation energy $E$ versus $\gamma$-ray energy $E_{\gamma}$ from particle-$\gamma$ coincidences recorded with the $^{232}$Th(d,p$\gamma)^{233}$Th reaction. The raw $\gamma$-ray spectra (a) are first unfolded (b) by the NaI response function. In the last step (c), the primary or first-generation $\gamma$-ray spectra are extracted as function of excitation energy $E$.}
 \label{fig:matrices}
 \end{center}
 \end{figure*}

In order to reduce the exposure of elastically scattered projectiles to the detectors, 
the 64 SiRi telescopes were placed in backward direction
covering eight angles from $\theta = 126^\circ$ to $140^\circ$
relative to the beam axis. The front  
and back detectors have thicknesses of $130$~$\mu$m and $1550$~$\mu$m, respectively. 
The CACTUS array consists of 28 collimated $5" \times 5"$ NaI(Tl) 
detectors with a total efficiency of $15.2$\% at $E_\gamma = 1.33$~MeV.

The particle-$\gamma$ coincidences with time information are sorted event by event. 
From the known charged-particle type and the kinematics of the reaction, the energies deposited in the telescopes 
can be translated to initial excitation energy $E$ in the residual nucleus. To avoid contamination from $\gamma$
 rays emitted by the fission fragments, we consider in this work only excitation energies that are well below the fission barriers of the studied actinides. In panel a) of Fig.~\ref{fig:matrices} is shown
the excitation energy versus $\gamma$ energy for the $^{232}$Th(d,p$\gamma)^{233}$Th reaction. For each energy bin $E$, 
the $\gamma$-spectra are unfolded using the procedure described in Ref.~\cite{gutt6}.
In this work we use newly determined NaI-response functions based on several in-beam $\gamma$ lines from excited
states in $^{56,57}$Fe, $^{28}$Si, $^{17}$O and $^{13}$C, where the relative efficiency with $\gamma$ energy could be extracted in a reliable way.  
The resulting matrix in Fig.~\ref{fig:matrices}b) describes the $\gamma$-ray energy distribution at each bin $E$
and is the starting point for the Oslo method.

An iterative subtraction technique has been developed to separate the distribution of first-generation (primary) $\gamma$ transitions from the total $\gamma$ cascade~\cite{Gut87}. Figure~\ref{fig:matrices}c) shows the final first-generation $\gamma$-ray matrix $P(E,E_{\gamma})$ for the $^{232}$Th(d,p$\gamma$)$^{233}$Th reaction. The subtraction technique is based on the assumption that the $\gamma$-decay spectra are the same whether the levels were initiated directly by the nuclear reaction or by $\gamma$ decay from higher-lying states. This assumption is automatically fulfilled when states have the same relative probability to be populated by the two processes, since $\gamma$-branching ratios are properties of the levels themselves. If the excitation bins contain many levels as for the actinides, it is likely to find the same $\gamma$-energy distribution independent of the method of population.

Fermi's golden rule predicts that the decay probability ($\lambda_{i \rightarrow f}$) may be factorized into the transition matrix element between the initial and final states, and the density at the final state~\cite{dirac,fermi}: 
\begin{equation}
\lambda_{i \rightarrow f} =   \frac{2 \pi}{\hbar} |{\langle}f |H^{\prime}|i {\rangle} | ^2  {\rho}_f.\
\label{eqn:1}
\end{equation}
Since the first-generation matrix $P(E,E_{\gamma})$ is proportional to the decay probability
to emit a $\gamma$-ray energy $E_{\gamma}$ from an initial excitation energy $E$, we may write the equivalent expression as: 
\begin{equation}
P(E, E_{\gamma}) \propto   {\cal{T}}_{i \rightarrow f}  \rho_f,\
\label{eqn:2}
\end{equation}
where ${\cal {T}}_{i \rightarrow f}$ is the $\gamma$-ray transmission coefficient, and $\rho_f = \rho (E -E_{\gamma})$ is the level density at the excitation energy after the primary $\gamma$-ray emission. According to the Brink hypothesis~\cite{brink}, the $\gamma$-ray transmission coefficient is approximately independent of excitation energy; only the transition energy $E_{\gamma}$ plays a role. Thus, we replace ${\cal {T}}_{i \rightarrow f}$ with ${\cal {T}}(E_{\gamma})$, giving
\begin{equation}
P(E, E_{\gamma}) \propto   {\cal{T}}(E_{\gamma}) \rho (E -E_{\gamma}).\
\label{eqn:3}
\end{equation}
This factorization allows a simultaneous extraction of level density and $\gamma$-ray transmission coefficient. 
In the next section, we will present the level densities for six actinide nuclei formed via eight reactions.

\section{Level densities}

The level densities obtained by fitting expression (\ref{eqn:3}) to the first-generation matrix determines only its functional form. It remains to normalize $\rho$ to data from other experimental results. At low excitation energy we use known levels to estimate the level density.  In conventional spectroscopy, a significant part of the levels are usually missing when the level density reaches 50 - 100 levels per MeV. Therefore, at high excitation energy the level density is normalized at the neutron separation energy $S_n$. The data point $\rho(S_n)$ is calculated from $\ell = 0$ neutron resonance spacings $D_0$  assuming a spin distribution~\cite{GC}
\begin{equation}
g(E=S_n,I) \simeq \frac{2I+1}{2\sigma^2}\exp\left[-(I+1/2)^2/2\sigma^2\right],
\label{eq:spindist}
\end{equation}
where $\sigma$ is the spin-cutoff parameter at the neutron separation energy.

Since the neutron resonance spacings only give the level densities for the lowest spins, e.g.~spin/parity $I^{\pi}=0^+$ and $1^+$ for $^{238}$U and $1/2^+$ for $^{239}$U, it is essential to know the spin distribution at $S_n$ in order to estimate the total level density. Calculations based on the combinatorial plus Hartree-Fock-Bogoliubov approach~\cite{goriely2008} indicate a spin-cutoff parameter for e.g.~$^{238}$U of $\sigma= 8.0-8.5$ at $S_n=6.154$ MeV. This value corresponds to a spin distribution for a nucleus exhibiting a rigid-body moment of inertia.

 \begin{figure}[]
 \begin{center}
 \includegraphics[clip,width=\columnwidth]{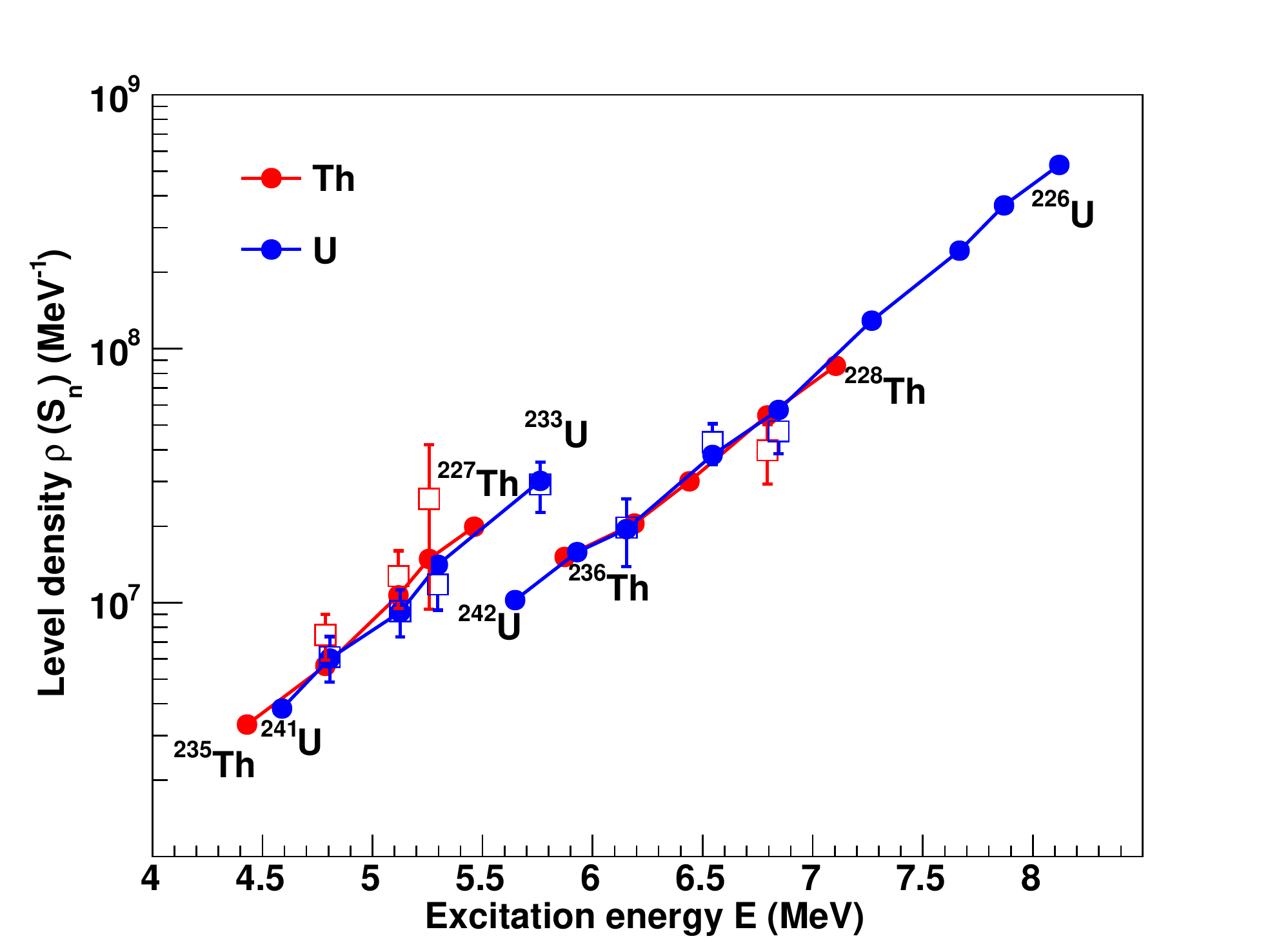}
 \caption{(Color online) Level densities at the neutron separation energies of Th and U predicted from global systematics~\cite{egidy2}. The experimentally obtained level densities deduced from neutron resonance spacings $D_0$ (open symbols with error bars) are shown for $^{229-233}$Th and $^{233-239}$U. The odd and even mass isotopes follow different curves.}
 \label{fig:rhobn}
 \end{center}
 \end{figure}

    In the global systematic study of level-density parameters by von Egidy and Bucurescu a rigid-body moment of inertia approach is used to determine the spin-cutoff parameter~\cite{egidy2}:
\begin{equation}
\sigma^2 = 0.0146 A^{5/3} \frac{1+\sqrt{1+4aU}}{2a},
\label{eqn:8}
\end{equation}
where $A$ is the mass number, $a$ is the level density parameter, $U=E-E_1$ is the intrinsic excitation energy, and $E_1$ is the back-shift parameter. In Table I the $a$ and $E_1$ parameters are taken from Ref.~\cite{egidy2}. It is satisfactory to see that the value of the spin-cutoff parameter $\sigma$ calculated from Eq.~(\ref{eqn:8}) is within the expected range of a rigid rotor. The last column of Table I shows the total level density calculated with experimental $D_0$ spacings from RIPL-3~\cite{RIPL3} and by means of the spin distribution of Eq.~(\ref{eq:spindist}).
 
 In the case of $^{232}$Th, the target nucleus $^{231}$Th is short-lived (25.52 h) and no neutron resonance spacing has been measured. Here, we base our estimate on the systematics~\cite{egidy2} obtained for the Th and U chain of isotopes as shown in Fig.~\ref{fig:rhobn}. Within the errors bars, the experimental level densities for 11  isotopes are well accounted for by the systematics, and we obtain an estimate for $^{232}$Th of $\rho(S_n)=(30\pm 8)\cdot 10^6$ MeV$^{-1}$.

  \begin{table}[htb]
    \caption{Parameters used to extract level densities at $S_n$ (see text).} 
    \begin{tabular}{c|c|ccccc}
    \hline
    \hline
    Nucleus   &$S_n$&   $a$    & $E_1$   &$\sigma(S_n)$&  $D_0$   &   $\rho(S_n)$ \\
              &(MeV)&(MeV$^{-1})$&(MeV)  &             &  (eV)    &(10$^6$ MeV$^{-1}$)\\
    \hline
    $^{231}$Th&5.118&   26.41  &  -0.42  &     7.78   & 9.6(15)  & 12.7(33)      \\
    $^{232}$Th&6.438&   25.87  & 0.30    &     8.05   &  -       & 30(8)$^a$     \\
    $^{233}$Th&4.786&   25.98  &  -0.58  &     7.82   &16.5(40)  &  7.4(15)      \\
    $^{237}$U &5.126&   25.60  &  -0.43  &     8.02   &14.0(10)  &  9.3(19)      \\
    $^{238}$U &6.154&   25.26  &   0.06  &     8.26   & 3.5(8)   & 20(6)         \\
    $^{239}$U &4.806&   26.67  &  -0.31  &     7.84   &20.3(6)   &  6.1(12)      \\

    \hline
    \hline
    \end{tabular}
    \\$^a$) Estimated from systematics~\cite{egidy2}, see Fig.~\ref{fig:rhobn}.
    \label{tab:parameters}
    \end{table}

\begin{figure}[b]
 \begin{center}
 \includegraphics[clip,width=\columnwidth]{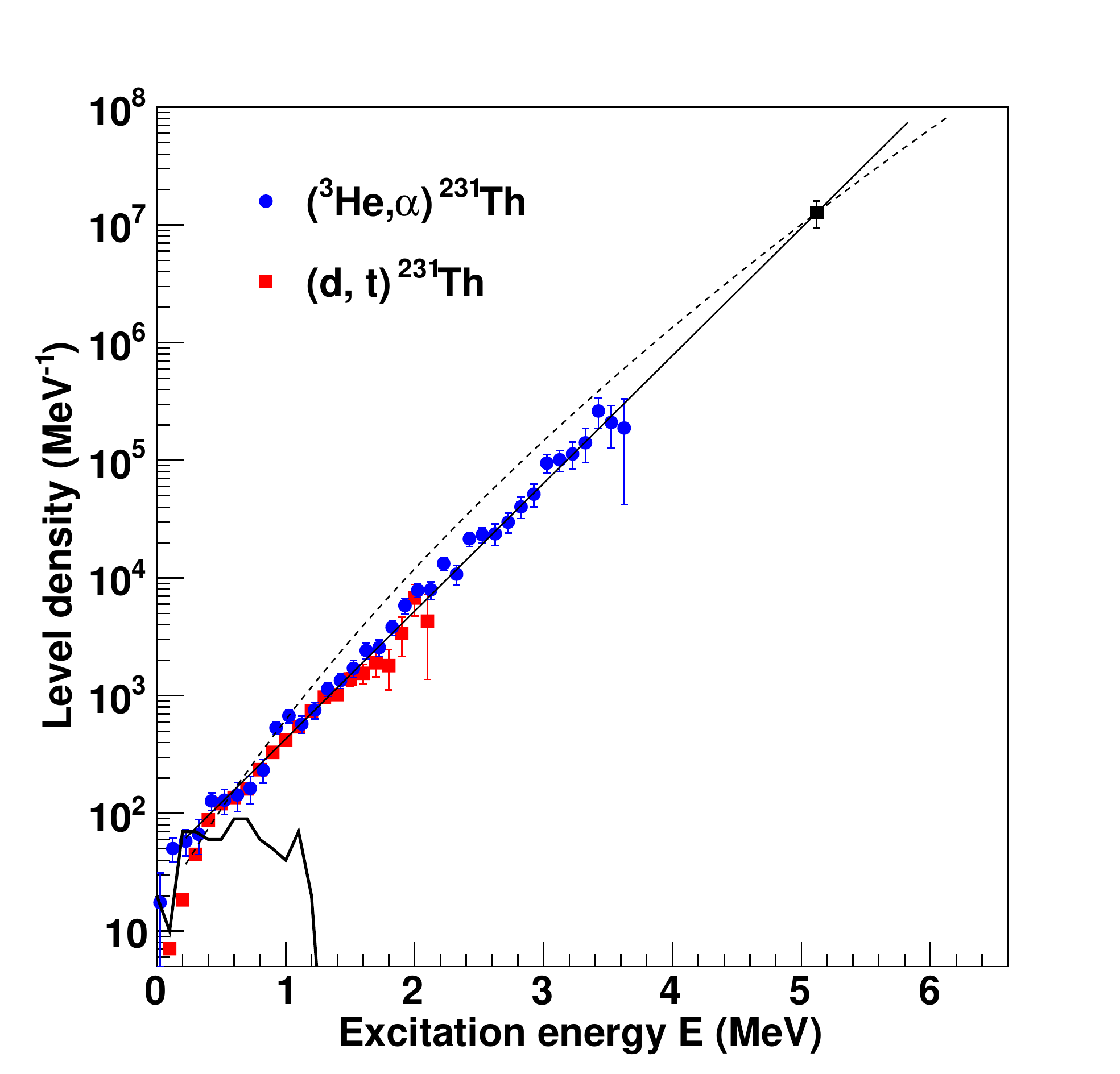}
 \caption{(Color online) Normalization of the nuclear level density of $^{231}$Th for the $(^3$He,$\alpha)$ and (d,t) reactions. At low excitation energies, the level density is normalized to known discrete levels (solid line). At higher excitation energies, the data are normalized to the constant-temperature level density with $T_{\rm CT}=0.40$ MeV (black line) going through  $\rho(S_n)$ (filled black square). For comparison, the Fermi-gas level density function with $a=26.41$ MeV$^{-1}$ and $E_1=-0.42$ MeV is shown (dashed line).}
 \label{fig:231Th}
 \end{center}
 \end{figure}

\begin{figure}[b]
 \begin{center}
 \includegraphics[clip,width=\columnwidth]{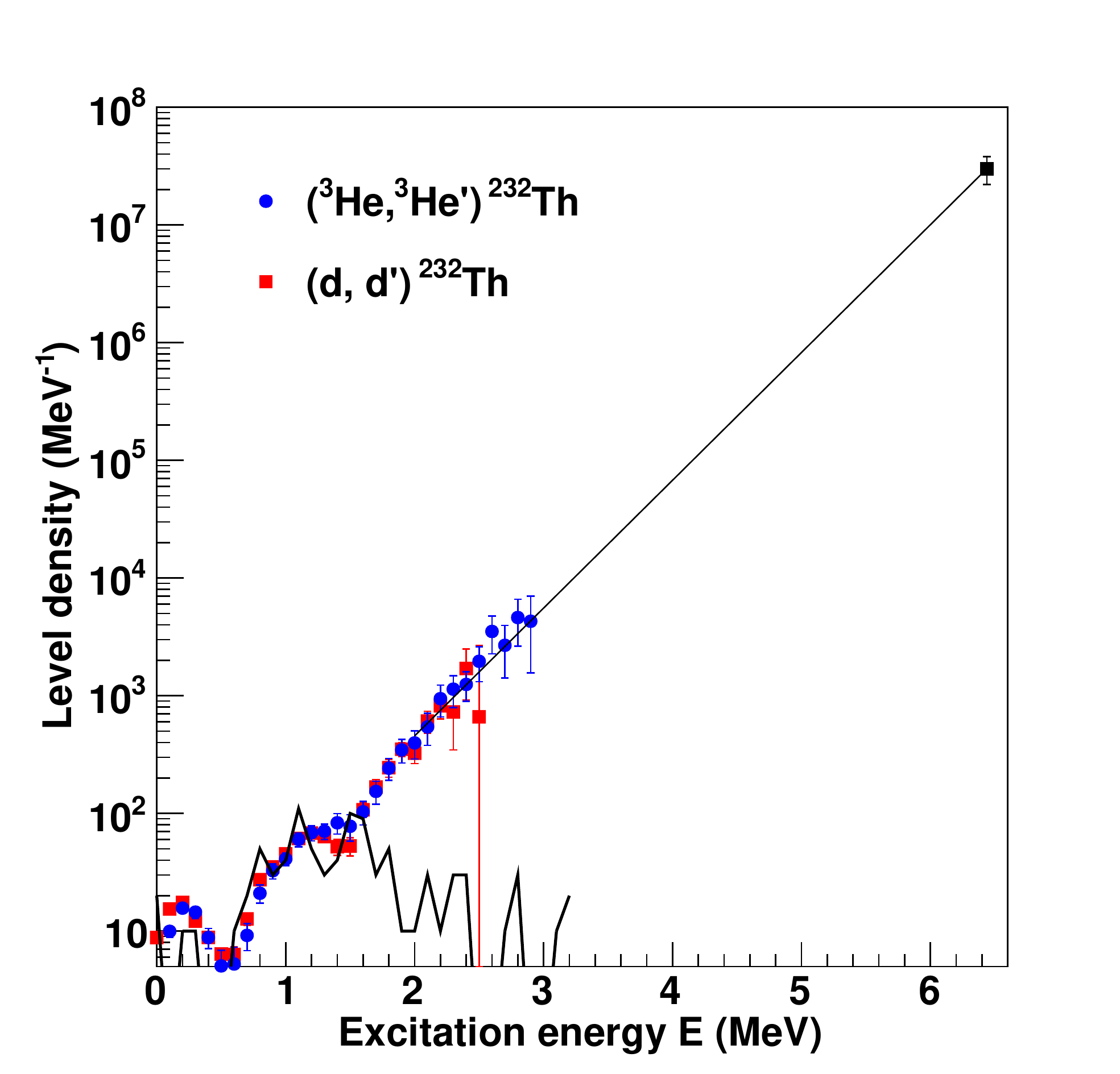}
 \caption{(Color online) Normalization of the nuclear level density of $^{232}$Th for the $(^3$He,$^3$He') and (d,d') reactions. See text of Fig.~\ref{fig:231Th}.}
 \label{fig:232Th}
 \end{center}
 \end{figure}

Figures \ref{fig:231Th} and \ref{fig:232Th} demonstrate the normalization procedure. We have also explored the ${^3}$He-induced reactions since the (d,x) reactions with 12 MeV beam energies only give data in a limited excitation-energy region. As a proof of principle, we compare here the level densities obtained from two different reactions giving the same residual nucleus. The reactions are $(^3$He,$\alpha)$ and (d,t) into $^{231}$Th and $(^3$He,$^3$He') and (d,d') into $^{232}$Th and give very similar results within the error bars.

At low-excitation energy we normalize to known discrete levels \cite{ENSDF} (see solid curve), which appear to be complete up to the excitation energy of $E\approx$ 0.2 MeV and 1 MeV in $^{231}$Th and $^{232}$Th, respectively. At the neutron separation energy we use the values of Table I. In order to normalize at the highest data points of our level densities, we  use the constant temperature formula~\cite{GC}
\begin{equation}
      \rho_{\rm CT}(E)=\frac{1}{T_{\rm CT}}\exp{\frac{E-E_0}{T_{\rm CT}}}
      \label{eq:ct}
\end{equation}
and extrapolate from $\rho(S_n)$ down to our data points\footnote{The temperature is expressed in units of MeV.}. In this formula, the parameter $T_{\rm CT}$ characterizes the slope of $\rho_{\rm CT}$ and $E_0$ the shift in excitation energy, determined by 
\begin{equation}
E_0=S_n-T_{\rm CT}\ln \left[\rho(S_n)T_{\rm CT}\right].
\end{equation}

The chosen constant temperature extrapolation is well justified in Fig.~\ref{fig:231Th}, where the experimental level density data points
follow closely a straight line in a log-plot. The alternative Fermi-gas formula (dashed line in Fig.~\ref{fig:231Th}) shows a convex shape in the log-plot that deviates significantly from the functional form of the experimental data. In fact, all the nuclei studied in this work follow the constant temperature formula, except $^{232}$Th in Fig.~\ref{fig:232Th}, where only the very last data points between $E=2 - 3$ MeV support this picture. However, there is no reason why this isotope should behave differently from its neighbors. The nuclei in this well deformed mass region have a uniform density of single particle orbitals, and no large nuclear-structure changes are expected. Thus, we use the same high-energy normalization method for all six isotopes by means of the constant temperature formula of Eq.~(\ref{eq:ct}).

It appears that in Figs.~\ref{fig:231Th} and \ref{fig:232Th} the extracted level densities are approximately independent of the specific light-ion reaction chosen, although we should keep in mind that we force the start and end of the level density curves to have the same value. However, the ($^3$He,$\alpha$) reaction is expected to populate a few $\hbar$ more spin compared to the (d,t) reaction. A closer look at the two level densities show that there is higher level density for the ($^3$He,$\alpha$) reaction around $E=1$ MeV that could be due to the decay to higher final spins. In spite of this, the functional form of the level density  appears to be approximately insensitive to spin effects and the reaction mechanism.

We also observe a reproduction of the level densities of the known low-energy levels. For the odd  $^{231}$Th case, the level density is high even for energies close to the ground state due to several Nilsson single-particle orbitals in the vicinity of the Fermi surface. In the even $^{232}$Th isotope the number of levels increases abruptly when the vibrational band heads appear in the 0.5 - 1.0 MeV region. At $E \approx 1.5$ MeV a second abrupt increase starts due to the breaking of nucleon Cooper pairs that requires roughly an energy of $2\Delta$, where $\Delta \approx 12/ \sqrt A$. In the following discussion, we consentrate on the results from the $^3$He induced reactions since these reactions gave best statistics and the largest $E$ range for $^{231,232}$Th.

\begin{figure}[t]
 \begin{center}
 \includegraphics[clip,width=\columnwidth]{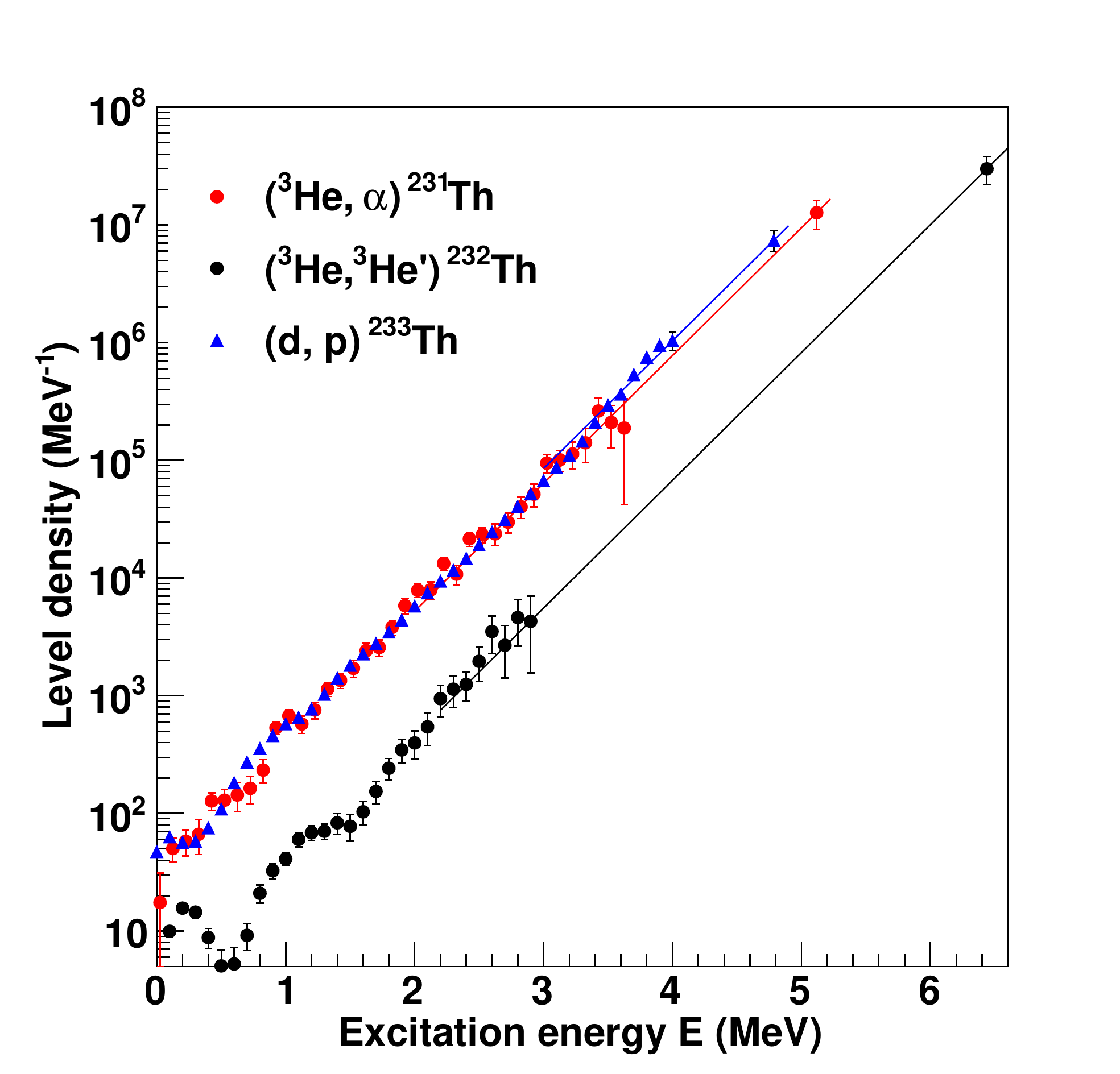}
 \caption{(Color online) Level densities for $^{231-233}$Th for the reactions with best statistics. The constant temperature extrapolations (solid lines) were calculated with $T_{\rm CT}=0.40$ MeV.}
 \label{fig:all_Th}
 \end{center}
 \end{figure}

\begin{figure}[t]
 \begin{center}
 \includegraphics[clip,width=\columnwidth]{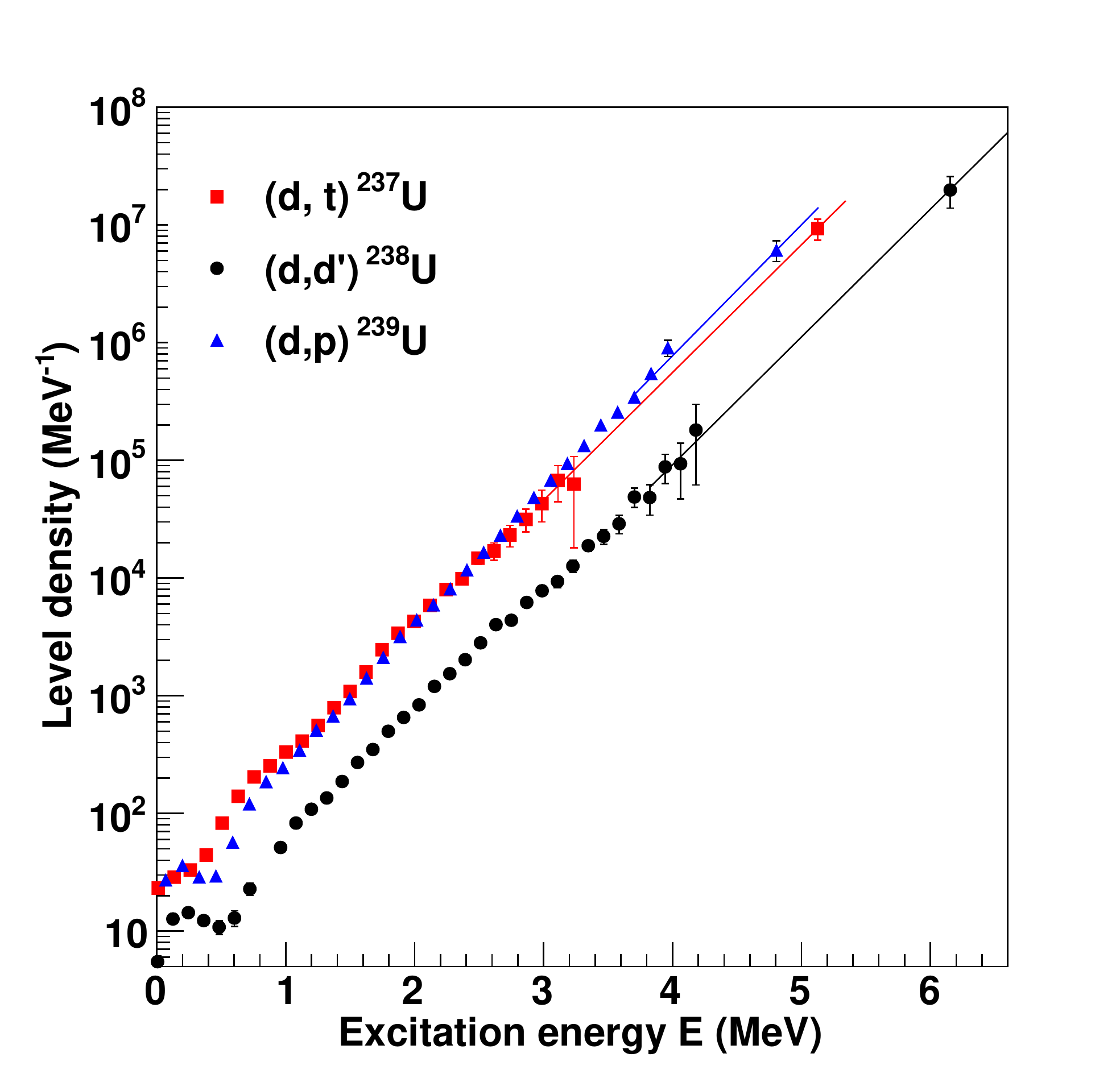}
 \caption{(Color online) Level densities for $^{237-239}$U. The constant temperature extrapolations (solid lines) were calculated with $T_{\rm CT}=0.40$ and $0.39$ MeV for $^{237,238}$U and $^{239}$U, respectively.}
 \label{fig:all_U}
 \end{center}
 \end{figure}

Figures~\ref{fig:all_Th} and \ref{fig:all_U} show the extracted level densities for $^{231-233}$Th and $^{237-239}$U, respectively. It is somewhat surprising that the even-even $^{238}$U does not show the same abrupt low-energy changes as seen for $^{232}$Th, even though the (d,d') reaction has the best particle energy resolution. The reason may be that the on-set of vibrational states and two-quasiparticle states overlap slightly in $^{238}$U giving a smoother increase in the level density.

One of the most striking properties of the six level densities is that they appear rather parallel in a log-plot. Furthermore, the odd-mass nuclei have a higher level density corresponding to a constant scaling factor. These features are dominantly determined by other experiments; ({\em i}) the number of discrete levels at low excitation energy and ({\em ii}) the total level densities estimated at $S_n$ by known neutron spacing data $D_0$. However, if the level density between these excitation energies were unknown, the conclusion of parallel level densities could not be drawn. It is therefore vital to know the full functional form.

The constant-temperature behavior is a puzzle. If the nucleus would have been in contact with a large heat bath, the concept of constant temperature (as in the canonical ensemble) would have been a reasonable result. However, such a heat bath is not present for isolated systems such as the nucleus. In Ref.~\cite{sumaryada}, pairing correlations and level densities are compared within the canonical, grand canonical, and microcanonical ensembles. All these models give a Fermi like level density and to our knowledge no fundamental and quantitative description of the observed constant-temperature behavior exists.

    A closer look to the data reveals structures in the measured level densities. 
    This is of great interest from a thermodynamic point of view and will be discussed in the next section.

\section{Thermodynamics}
A challenging goal in nuclear physics is to trace thermodynamic quantities as function of excitation energy. These average quantities represent the only observables for systems where the levels are so close and numerous that it is impossible to measure them separately. With the present technique, we are able to almost bridge the gap between low-lying states and the level density at $S_n$.

The density of levels as a function of excitation energy is the starting point to extract quantities such as entropy, temperature and heat capacity. The isolated atomic nucleus is a perfect system for the microcanonical ensemble theory. It has a sharp excitation energy and the number of particles is fixed by $N$ and $Z$. Furthermore, the high incompressibility justifies the assumption of a constant volume for the modest excitation energy region considered here. Despite this ideal ensemble, the statistical properties of the nucleus are difficult to describe theoretically.

The level density $\rho(E)$ is proportional to the number of states accessible to the nuclear system at a given excitation energy $E$. We may define the multiplicity of states as $\Omega(E)=\rho/\rho_0$, where $\rho_0$ is the level density close to the ground state in the even-even isotope. Thus, we assure that the multiplicity becomes $\Omega=1$ at the ground state. The entropy in the microcanonical ensemble is given by
\begin{equation}
S(E)= k_B \ln \Omega (E),
\end{equation}
where $k_B$ is the Boltzmann constant. 

\begin{figure}[t]
 \begin{center}
 \includegraphics[clip,width=\columnwidth]{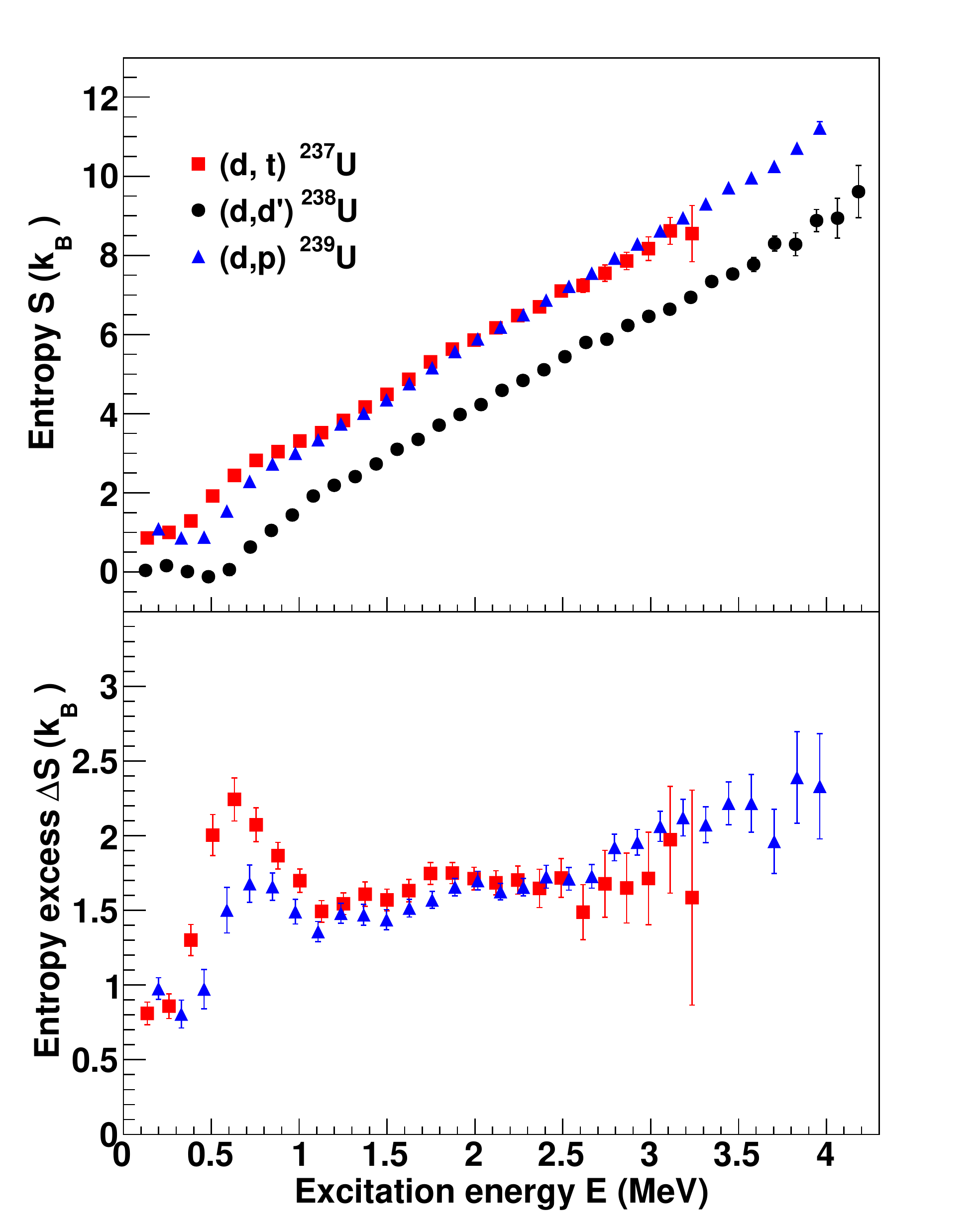}
 \caption{(Color online) Entropies for $^{237-239}$U (upper panel) and entropy excess of $^{237,239}$U compared to $^{238}$U (lower panel).}
 \label{fig:entropy3}
 \end{center}
 \end{figure}

Since the uraniums represent the most complete data set (Fig.~\ref{fig:all_U}) we will focus on these isotopes in this section. However, the results also hold for the thoriums.
In Fig.~\ref{fig:entropy3} the entropies for $^{237-239}$U are displayed. The odd-mass isotopes have similar entropies with slightly different constant-temperatures of $T_{\rm CT}=$ 0.40(1) and 0.39(1) MeV for $^{237}$U and $^{239}$U, respectively. In contrast to the even-mass  $^{238}$U isotopes, their entropies at low excitation energy are smeared out. This is partly due to the fact that single neutron band heads appear at various excitation energies and partly due to blocking effects from the last neutron. In the lower panel the difference in entropy $\Delta S = S({\rm odd-mass}) - S({\rm even-even})$ is evaluated. In the excitation region of $E = 1-3$ MeV, the excess of entropy stabilizes around $\Delta S = 1.6(2) k_B$. This corresponds to 5 times higher level densities due to the unpaired last neutron. For the thoriums, despite the poor data on $^{232}$Th, the entropy difference reaches a much higher value of $\Delta S \approx 2.3 k_B$. This indicates that the valence neutron in thorium has more available orbitals close to the Fermi surface than the uraniums. It is important to realize that this valence neutron is not placed in a specific orbital with given spin and parity, but has the average property of all valence neutron orbits at a certain excitation bin.
  
In the microcanonical ensemble the temperature and heat capacity can be expressed by
\begin{equation}
T(E)=(\partial S/\partial E)^{-1}
\label{eqn:t}
\end{equation}
and 
\begin{equation}
C_V(E)=(\partial T/\partial E)^{-1},
\label{eqn:cv}
\end{equation}
respectively. Small statistical deviations in the entropy may give rise to large contributions in the temperature and heat capacity. In order to reduce these fluctuations, the differentiation of $S$ is performed by a least square fit of a straight line to five adjacent data points at a time. The result is an  effective smoothing of about 0.6 MeV, which is larger than the energy resolution of the experimental level density and may reduce the information content of the data. On the other hand, we are here looking for changes due to pairing effects that is expected to be of the order of 2$\Delta$.

 \begin{figure}[t]
 \begin{center}
 \includegraphics[clip,width=\columnwidth]{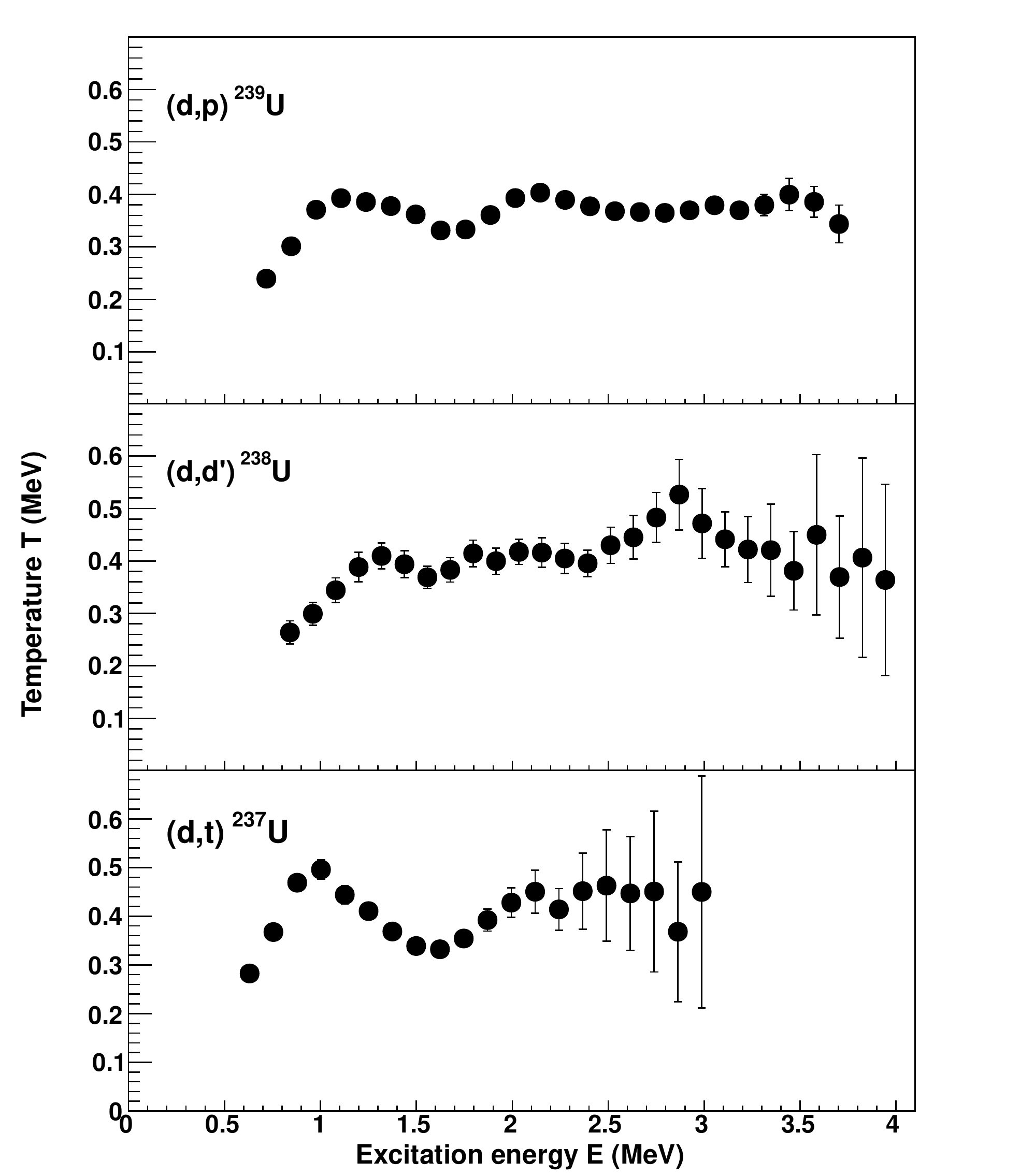}
 \caption{Microcanonical temperatures for $^{237-239}$U as function of excitation energy from this work.}
 \label{fig:temp3}
 \end{center}
 \end{figure}
 \begin{figure}[t]
 \begin{center}
 \includegraphics[clip,width=\columnwidth]{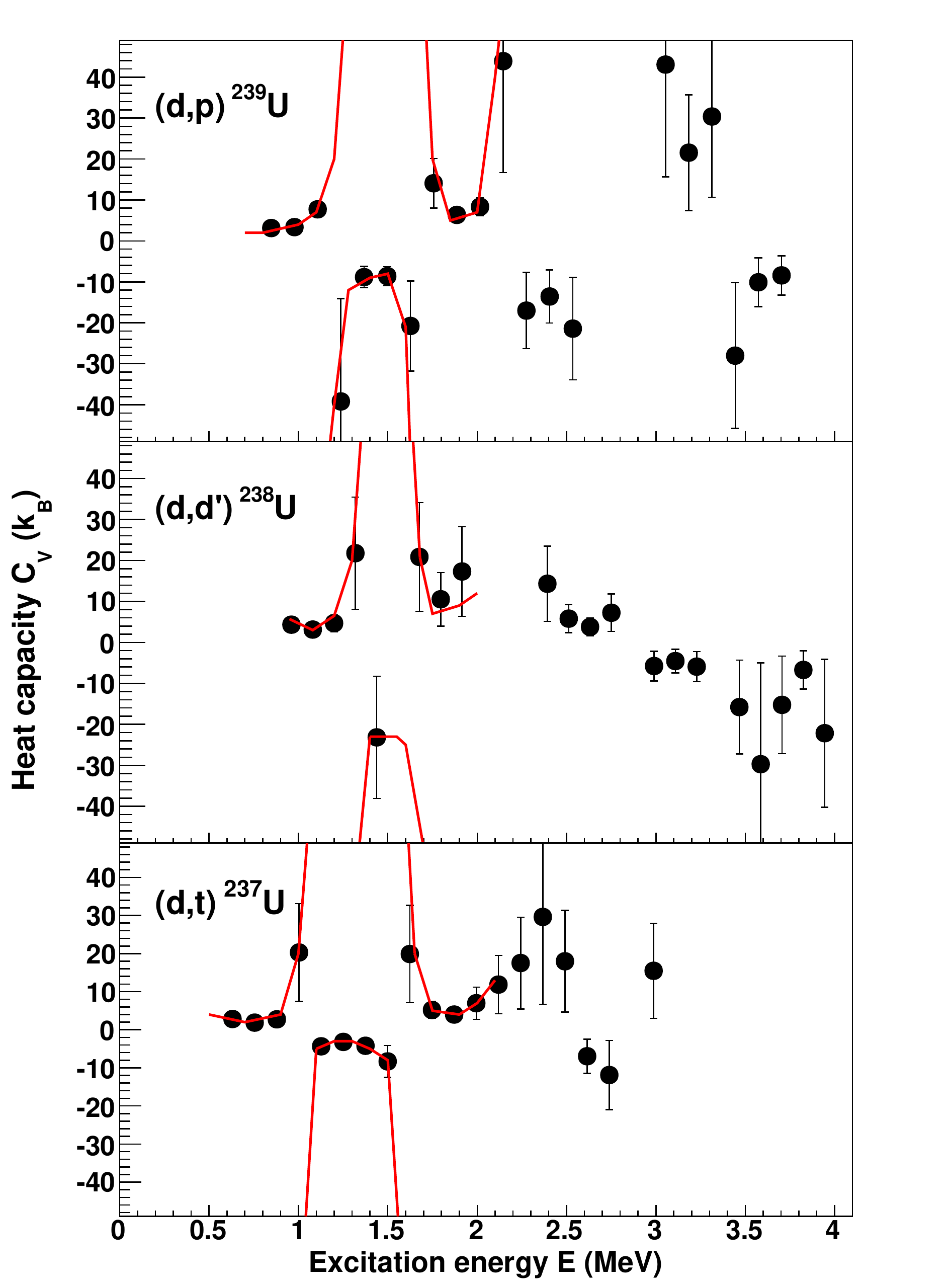}
 \caption{(Color online) Microcanonical heat capacity as function of excitation energy for $^{237-239}$U. There are several
 data points out of scale. The red curves are drawn to guide the eye.}
 \label{fig:heatcv3}
 \end{center}
 \end{figure}

The caloric curves $T(E)$ for $^{237-239}$U are shown in
Fig.~\ref{fig:temp3}. The flat entropy in the ground band of $^{238}$U gives
$\partial S/\partial E\approx 0$, which means that the temperature is undefined below $E \approx 0.6$ MeV. In fact, below this excitation energy, the few levels and the onset of rotation and vibrations make the concept of temperature difficult to adopt.  Above this energy all 
three isotopes display an increasing temperature up to $E \approx$ 1.0 - 1.2  MeV with a subsequent drop to a minimum temperature 
at $E \approx$ 1.7  MeV. This drop in temperature with increasing energy is probably due to the breaking of the first Cooper pair. 
This is not a specific pair defined by spin and parity, but an average of all types of broken pairs in that energy bin. 
At $E\approx $ 1.4 - 1.5 MeV this transition releases energy to the isolated system from the latent pairing energy. 
From the temperature valley at $E\approx $ 1.7 MeV the temperature increases with energy and then exhibits 
a constant temperature of $T\approx$ 0.4 MeV for all three isotopes. 

A large level density is mainly built from all possible combinations of available single nucleons, which add up to a given excitation energy. However, the nucleons interact strongly with each other through the pairing force and the Pauli principle plays an important role. The pairing correlations are expected to quench as energy and number of quasiparticles increases. This gives rise to faster breaking of Cooper pairs with excitation energy and thus giving a boost in level density;  each broken pair contributes with $\exp(2\Delta S)$ times more levels. These two components of the level density may balance each other in such a way that the level density takes the simple exponential form. A closer look of the caloric curve in Fig.~\ref{fig:temp3} indeed reveals oscillations around the constant-temperature value.

The fact that $T(E)$ actually drops at certain excitation energies is not in accordance with every-day experience. If we add energy to a system, it gets warmer. We are also familiar with the fact that melting ice keeps the temperature constant at 0$^{\circ}$C. However, to observe that a system gets colder by adding energy is quite extraordinary. 

When the temperature decreases with increasing energy, the phenomenon of negative heat capacity occurs.
Negative heat capacity  has for a long time been known for certain systems. Some stars and star clusters cool down when energy is added~\cite{thirring,lynden} and small objects like atom clusters display the same feature~\cite{bixon,labastie,schmidt}. In Au+Au multifragmentation experiments negative heat capacity has been seen~\cite{agostino}, and the Oslo group observed the same for heated $^{166,167}$Er~\cite{melby}.

Figure~\ref{fig:heatcv3} shows the heat capacity as function of excitation energy for $^{237-239}$U. From Eq.~(\ref{eqn:cv}) we see that $C_V$ is undefined ($\pm \infty$) when the temperature is constant with excitation energy. The occurrence of such poles is exactly what happens at several places in the $C_V(E)$ curves of  Fig.~\ref{fig:heatcv3} making the data points rather chaotic. However, for low E the curves has a clear physical message, and we have drawn red curves to guide the eye for the discussion.

In the following we concentrate on the clearest case, namely $^{237}$U. At the lowest energies, $C_V$ increases to $+\infty$ at $E=1$ MeV. Then $C_V$ switches to  $-\infty$ and increases to $\approx -5 k_B$, before it approaches $-\infty$ at $E=1.5$ MeV. Then $C_V$ goes to $+\infty$ and decreases down to a level of $\approx +5 k_B$ at $E=1.7$ MeV. It is easy to recognize stage by stage this behavior of $C_V$ from the caloric curve of $^{237}$U in Fig.~\ref{fig:temp3}. We find that the process of breaking the first Cooper pairs in $^{237}$U takes place between $E=1.0$ and $1.6$ MeV with a corresponding cooling from $T=0.50$ to $0.34$ MeV.

From the $T(E)$ and $C_V(E)$ curves, we may get a hint of the mechanisms behind the constant-temperature level density functional forms. 
Both $T$ and $C_V$ indicate that unknown processes start to contribute  as energy
increases. Clearly seen in Fig.~\ref{fig:temp3} are sudden decreases in the caloric curve,
which is indicative of underlying mechanisms reducing the temperature of the
system. This could be interpreted as a continuous melting of Cooper pairs throughout the energy region studied here.

\section{Conclusions}
\label{sec:con}

The level densities of $^{231-233}$Th and $^{237-239}$U have been determined using the Oslo method. Similar functional forms of the level density have been extracted with different nuclear reactions leading to the same residual nucleus. This consistency gives confidence to the assumptions behind the method. The level densities of all six isotopes exhibit a constant temperature level density behavior with $T_{\rm CT} \approx 0.40$ MeV when normalized to known anchor points. There is a clear increase in level density for the odd-mass Th and U isotopes compared to even-mass isotopes. The corresponding excess in entropy $\Delta S$ reveals the degree of freedom for the average valence neutron outside the even-even core.

Negative heat capacity is a fingerprint for a phase transition. We observe several poles in the $C_V(E)$ curve, making the assumption of an almost continuous melting of Cooper pair plausible. However, there is a great need for a proper theoretical description of the constant-temperature shape of the level density as well as the rich thermodynamics found  in the actinides.

\acknowledgements
We would like to thank J.~M{\"{u}}ller, E.A.~Olsen, A.~Semchenkov and J.~Wikne at the Oslo Cyclotron Laboratory for providing the stable and high-quality deuterium and 3He beams during the experiment, the Lawrence Livermore National Laboratory for providing the $^{232}$Th target and the GSI Target Laboratory for the production of the $^{238}$U target. This work was supported by the Research Council of Norway (NFR), the French national research programme GEDEPEON, the US Department of Energy under Contract No.~DE-AC52-07NA27344, the National Research Foundation of South Africa, the European Commission within the 7th Framework Programme through Fission-2010-ERINDA (Project No.~269499) and by the European Atomic Energy CommunityÕs 7th Framework Programme under grant agreement no.~FP7-249671 (ANDES).

\vfill
\end{document}